# Higher-order optical-polarization

**Ravi S.Singh[1,*] and Hari Prakash[2]**

[1]*Department of Physics, D.D.U. Gorakhpur University, Gorakhpur-273009 (U. P.)-INDIA*
[2]*Department of Physics, University of Allahabad, Allahabad-211002 (U. P.)-INDIA*
[*]*Corresponding author: yesora27@gmail.com*

Polarization properties of optical field in Schrödinger Cat states or Even- or Odd-Coherent states are investigated by generalizing usual concept of optical-polarization to render the concept of Higher-order optical-polarization. It is observed that Higher-order optical-polarization is a basis-dependent property of optical field.

Polarization of Light, a centuries-old fundamental concept, ensures transversal nature of optical field. While Stokes parameters characterize states of optical-polarization completely in Classical Optics [1], they attain the roles of operators in Quantum Optics. Their expectation values can be discrete or continuous depending upon the states of light and, thus, furnishing descriptions in single-photonic and continuous-variable domain [2-5]. Theory of optical-polarization is generally dwelt with two extreme cases: Perfect polarized light and Unpolarized light. In between these extremities there exist infinitely many partial optical-polarized states.

A first rigorous study of unpolarized light in quantum domain is carried out by Prakash and Chandra [6], and independently by Agarwal [7], who discovered the structure of its density operator. These authors [6] demonstrated that the characterization of optical-polarization by Stokes parameters is inadequate as they embody only linear interaction between optical field-modes. Recently, insufficiency of these parameters is observed in Ref. [8]. Mehta and Sharma [9] defined perfect polarized state of a plane monochromatic light by demanding vanishing amplitude in one linearly polarized mode after passing through a compensator followed by a rotator. Such a polarized optical field is truly a single-mode field (having no signal in one orthogonal mode) described statistically by single random complex amplitude (CA). A criterion [10] for perfect optical-polarization state with the help of 'index of polarization' (IOP) is set up by the ratio of CAs in two transverse orthogonal bases-modes of which non-random values provide 'ratio of real amplitudes' and 'difference in phases', characteristic parameters for polarized light [11].

While notable idea [12], in terms of probability distribution functions on Poincare sphere, for characterization of optical-polarization has been proposed by tracing analogies between Stokes parameters with components of spin-angular momenta, it is critically emphasized [13, 14] that the Hilbert space spanned by optically-polarized filed-states is not alike to that by states of Jordon-Schwinger spin-angular momenta [15]. This dissimilarity induces an alternative quasi-probability distribution function for optical-polarization [16, 17]. One may, therefore, sense that rigorous description of perfect optical-polarization and its generalization offers a moot question.

In this letter we introduce, classically, Higher ($n^{th}$) order Optical-polarization (HOP) in the 'basis of description' if, for a single mode optical field, ratio of CAs in transverse orthogonal modes is random but all its multiple powers of some positive integers, say, 'n' are non-random parameters. Higher-order optical-polarized beam is not a single mode optical field because, although it maintains non-random value for ratio of real amplitudes, phase may take equally probable values among 'n' non-random values in steps of $2\pi/n$. The usual concept of optical-polarization results when n = 1. Quantum version of the definition is provided by involving quantum CAs (Bosonic annihilation operators).

A plane monochromatic unpolarized optical field propagating along z-direction can, in general, be described by vector potential, $\boldsymbol{\mathcal{A}}$ ,

$$\boldsymbol{\mathcal{A}} = (\hat{\mathbf{e}}_x\mathcal{A}_x + \hat{\mathbf{e}}_y\mathcal{A}_y)e^{-i\psi}\,;\ \mathcal{A}_{x,y} = \underline{A}_{x,y}\,e^{-i\psi},$$
$$\underline{A}_{x,y} = A_{0x,0y}\,e^{i\phi_{x,y}};\ \psi = \omega t - kz, \qquad (1)$$

where $\mathcal{A}_{x,y}$ are defined as magnitudes of analytic signal $\boldsymbol{\mathcal{A}}$, $\underline{A}_{x,y}$ as classical CAs, $A_{0x,0y}$ as real amplitudes possessing, in general, random spatio-temporal variation with angular frequency, $\omega$, $\phi_{x,y}$ (phase parameters) may take equally probable random values in the interval $0 \le \phi_{x,y} < 2\pi$ in linear polarization-basis $(\hat{\mathbf{e}}_x, \hat{\mathbf{e}}_y)$, $\mathbf{k}$ $(= k\hat{\mathbf{e}}_z)$ is propagation vector of magnitude k, and $\hat{\mathbf{e}}_{x,y,z}$ are unit vectors along respective x-, y-, z-axes forming right handed triad. Light may be said to be polarized only when the ratio of CAs must keep non-random values, i.e., $\underline{A}_y/\underline{A}_x = p$, a non-random complex parameter defining 'index of polarization' (IOP). Clearly, polarized optical field (p is known) is a single mode field as only one random CA suffices for its complete statistical description (other orthogonal CA is specified by p). If one introduces new parameters $A_0$ (real random amplitude), $\chi_0$ (polar angle), $\Delta_0$ (azimuth angle), $\phi$ (random phase) on a Poincare sphere, satisfying inequalities $0 \le A_0$, $0 \le \chi_0 \le \pi$, $-\pi < \Delta_0 \le \pi$, $0 \le \phi < 2\pi$, respectively, and preserving transforming equations in terms of old parameters, $A_0 = (A_{0x}^2 + A_{0y}^2)^{1/2}$; $\chi_0 = 2\tan^{-1} A_{0y}/A_{0x})$ and $\Delta_0 = \phi_y - \phi_x$; $\phi = (\phi_x + \phi_y)/2$, the analytic signal, $\boldsymbol{\mathcal{A}}$, Eq. (1), finds an instructive compact form

$$\boldsymbol{\mathcal{A}} = \hat{\boldsymbol{\varepsilon}}_0 \mathcal{A};\ \mathcal{A} = \underline{A} e^{-i\Psi};\ \underline{A} = A_0\, e^{i\phi},$$
$$\hat{\boldsymbol{\varepsilon}}_0 = \hat{\mathbf{e}}_x \cos\frac{\chi_0}{2} e^{-\Delta_0/2} + \hat{\mathbf{e}}_y \sin\frac{\chi_0}{2} e^{-\Delta_0/2}. \qquad (2)$$

The Eq. (2) may be interpreted as the single mode polarized optical field, statistically described by single CA, $\underline{A}$ and polarized in the fixed direction, $\hat{\boldsymbol{\varepsilon}}_0$ determined by non-random angle parameters $\chi_0$ and $\Delta_0$ in the Poincare sphere specifying the mode, $(\hat{\boldsymbol{\varepsilon}}_0, \mathbf{k})$. The complex vector $\hat{\boldsymbol{\varepsilon}}_0$ is a unit vector ($\boldsymbol{\varepsilon}_0^*.\boldsymbol{\varepsilon}_0 = 1$) giving expression of IOP, $p = \tan\frac{\chi_0}{2} e^{i\Delta_0}$. Obviously, the state of polarization of light is specified by the non-random values of p, which, in turn, is fixed by non-random values of $\chi_0$ and $\Delta_0$ defining a point ($\hat{\boldsymbol{\varepsilon}}$), in the unit Poincare sphere, similar to Stokes parameters. All typical parameters in ellipsometry such as major axis, minor axis and orientation angles of the polarization-ellipse can be determined if p of optical field and one CA are specified.

In elliptic-polarization basis ($\hat{\boldsymbol{\varepsilon}}$, $\hat{\boldsymbol{\varepsilon}}_\perp$) [18] such polarized light (Eq. 2) retain IOP, $p_{(\hat{\varepsilon},\hat{\varepsilon}_\perp)}$, another non-random parameter showing vivid dependence on $\hat{\boldsymbol{\varepsilon}}_0$,

$$p_{(\hat{\varepsilon},\hat{\varepsilon}_\perp)} = \underline{A}_{\hat{\varepsilon}_\perp}/\underline{A}_{\hat{\varepsilon}} == \frac{\hat{\boldsymbol{\varepsilon}}_\perp^*.\hat{\boldsymbol{\varepsilon}}_0}{\hat{\boldsymbol{\varepsilon}}^*.\hat{\boldsymbol{\varepsilon}}_0}, \qquad (3)$$

where $\hat{\boldsymbol{\varepsilon}}_\perp$ is orthogonal complex unit vector ( $\hat{\boldsymbol{\varepsilon}}_\perp^*.\hat{\boldsymbol{\varepsilon}}_\perp = 1$, $\hat{\boldsymbol{\varepsilon}}_\perp^*.\hat{\boldsymbol{\varepsilon}} = 0$). This formula caters interchangeability of IOP's between bases of descriptions. Parametrizing CAs $\underline{A}_{\hat{\varepsilon}}$ and $\underline{A}_{\hat{\varepsilon}_\perp}$ by $\underline{A}_{\hat{\varepsilon}} = A_{0\hat{\varepsilon}} \exp(i\phi_{\hat{\varepsilon}})$, $\underline{A}_{\hat{\varepsilon}_\perp} = A_{0\hat{\varepsilon}_\perp}\exp(i\phi_{\hat{\varepsilon}_\perp})$, with real-amplitudes $(A_{0\hat{\varepsilon}}, A_{0\hat{\varepsilon}_\perp})$ and phase-parameters $(\phi_{\hat{\varepsilon}}, \phi_{\hat{\varepsilon}_\perp})$, the eq. (3) reveals that such a polarized beam maintain non-random values of (i) 'ratio of real amplitudes', $A_{0\hat{\varepsilon}_\perp}/A_{0\hat{\varepsilon}}$, and (ii) 'difference in phase', $(\phi_{\hat{\varepsilon}_\perp} - \phi_{\hat{\varepsilon}})$, in ($\hat{\boldsymbol{\varepsilon}}$, $\hat{\boldsymbol{\varepsilon}}_\perp$). The preceding discussions highlight the fact that the state of polarized optical field (Eq. 2) may be described by considering any basis of description.

The concept of HOP may be propounded by demanding non-random values of all multiple powers of positive integer, say, n of ratio of CAs except n = 1. The IOP for light in n[th] -order HOP read as

$$p_{(\hat{\varepsilon},\hat{\varepsilon}_\perp),n} \equiv (\underline{A}_{\hat{\varepsilon}_\perp}/\underline{A}_{\hat{\varepsilon}})^n. \qquad (4)$$

The positive integer, n takes unit value for usual concept of optical-polarization (cf. eq. 3). For HOP, where n is greater than unity, the random value of $\underline{A}_{\hat{\varepsilon}_\perp}/\underline{A}_{\hat{\varepsilon}}$ and non-random value of

$$(\underline{A}_{\hat{\varepsilon}_\perp}/\underline{A}_{\hat{\varepsilon}})^n = p_{(\hat{\varepsilon},\hat{\varepsilon}_\perp),n} = \left|p_{(\hat{\varepsilon},\hat{\varepsilon}_\perp),n}\right| \exp(i\Delta_{(\hat{\varepsilon},\hat{\varepsilon}_\perp),n}) \qquad (5)$$

communicate us that while ratio of real amplitudes, $A_{0\hat{\varepsilon}_\perp}/A_{0\hat{\varepsilon}}$ has a non-random value, $\left|p_{(\hat{\varepsilon},\hat{\varepsilon}_\perp),n}\right|^{1/n}$, difference in phase, $\phi_{\hat{\varepsilon}_\perp}$- $\phi_{\hat{\varepsilon}}$ may have equally probable values among the n non-random values, $\frac{1}{n}\left(\Delta_{(\hat{\varepsilon},\hat{\varepsilon}_\perp),n} + 2r\pi\right)$ with r = 0, 1, 2,...(n-1), in steps of $(2\pi/n)$. Obviously this imprecise knowledge of phases makes HOP not to be characterized by single CA and is, therefore, not a single-mode optical field contrary to the usual perfect polarized light.

In Quantum Optics the optical field, Eq. (1) is described by operatic-version of vector potential operator,

$$\hat{\boldsymbol{\mathcal{A}}} = \left(\frac{2\pi}{\omega V}\right)^{1/2} [(\hat{\mathbf{e}}_x \hat{a}_x + \hat{\mathbf{e}}_y \hat{a}_y)e^{-i\Psi} + \text{h.c.}],$$
$$= \left(\frac{2\pi}{\omega V}\right)^{1/2} [(\hat{\boldsymbol{\varepsilon}}\, \hat{a}_{\hat{\varepsilon}} + \hat{\boldsymbol{\varepsilon}}_\perp \hat{a}_{\hat{\varepsilon}_\perp})e^{-i\Psi} + \text{h.c.}], \qquad (6)$$

in linear-polarization basis $(\hat{\mathbf{e}}_x, \hat{\mathbf{e}}_y)$ or in elliptic-polarization basis $(\hat{\boldsymbol{\varepsilon}}, \hat{\boldsymbol{\varepsilon}}_\perp)$, respectively, where ω is angular frequency of the optical field and V is the quantization volume of the cavity, h.c. stands for Hermitian conjugate. Using orthonormal properties of $\hat{\boldsymbol{\varepsilon}} (= \varepsilon_x \hat{\mathbf{e}}_x + \varepsilon_y \hat{\mathbf{e}}_y)$ and $\hat{\boldsymbol{\varepsilon}}_\perp (= \varepsilon_{\perp x}\hat{\mathbf{e}}_x + \varepsilon_{\perp y}\hat{\mathbf{e}}_y)$ the annihilation operators $\hat{a}_{\hat{\varepsilon}}$ ($\hat{a}_{\hat{\varepsilon}_\perp}$) are related with those in linear-polarization basis $(\hat{\mathbf{e}}_x, \hat{\mathbf{e}}_y)$ by,

$$\hat{a}_{\hat{\varepsilon}} = \varepsilon_x^* \hat{a}_x + \varepsilon_y^* \hat{a}_y,\ \hat{a}_{\hat{\varepsilon}_\perp} = \varepsilon_{\perp x}^* \hat{a}_x + \varepsilon_{\perp y}^* \hat{a}_y, \qquad (7)$$

satisfying usual Bosonic-commutation relations.

The dynamical pure state of a monochromatic optical beam, propagating along z-axis and polarized in the mode $(\hat{\boldsymbol{\varepsilon}}_0, \mathbf{k})$, may be specified by a ket vector $|\psi\rangle$ in Hilbert space. Evidently, such light beam doesn't have signal (photons) in orthogonal mode $(\hat{\boldsymbol{\varepsilon}}_{0\perp}, \mathbf{k})$ which manifests into the expression,

$$a_{\hat{\varepsilon}_{0\perp}}|\psi\rangle = 0, \qquad (8)$$

yielding, on applying Eq.(7), $(\varepsilon_{0\perp x}^* \hat{a}_x + \varepsilon_{0\perp y}^* \hat{a}_y)|\psi\rangle = 0$. The orthogonality relation between $\hat{\boldsymbol{\varepsilon}}_0$ and $\hat{\boldsymbol{\varepsilon}}_{0\perp}$ manipulates above equation as,

$$\hat{a}_y |\psi\rangle = p\hat{a}_x|\psi\rangle, \qquad (9)$$

giving $p = \epsilon_{0y}/\epsilon_{0x}$ $(= \tan\frac{\chi_0}{2} e^{i\Delta_0}$, from Eq. 2$)$ for perfect polarized light in Quantum Optics [19]. Eq. (9) may, therefore, be treated as quantum-analogue of $\underline{A}_y = p\,\underline{A}_x$. The defining equation, Eq. (9) for perfect optical-polarization may be casted in elliptic-polarization basis $(\hat{\boldsymbol{\varepsilon}}, \hat{\boldsymbol{\varepsilon}}_\perp)$ as

$$\hat{a}_{\hat{\varepsilon}_\perp}|\psi\rangle = p_{(\hat{\varepsilon},\hat{\varepsilon}_\perp)} a_{\hat{\varepsilon}}|\psi\rangle, \tag{10}$$

upon which repetitive applications by annihilation operator 'r'-times (r, any positive integer) yields,

$$(\hat{a}_{\hat{\varepsilon}_\perp})^r|\psi\rangle = p_{(\hat{\varepsilon},\hat{\varepsilon}_\perp),r} (a_{\hat{\varepsilon}})^r|\psi\rangle. \tag{11}$$

We establish the quantum version of the criterion (see eq. 4) for HOP by demanding,

$$(\hat{a}_{\hat{\varepsilon}_\perp})^n|\psi\rangle = p_{(\hat{\varepsilon},\hat{\varepsilon}_\perp),n} (a_{\hat{\varepsilon}})^n|\psi\rangle, \tag{12}$$

for smallest value n of a positive integer, r. It is evident that for n = 1, the usual definition of optical-polarization results (cf. Eq. 10) revealing that ordinary optical-polarization is first-order HOP. For n > 1, Eq. (9) is fulfilled only when n is a multiple of positive integer r. For mixed state of optical field specified by the density operator, ρ, defining prescription, Eq. (12) assumes the form,

$$(\hat{a}_{\hat{\varepsilon}_\perp})^n\rho = p_{(\hat{\varepsilon},\hat{\varepsilon}_\perp),n} (a_{\hat{\varepsilon}})^n\rho\,. \tag{13}$$

Assuming optical-field in bimodal coherent states, $|\alpha,\beta\rangle_{(\hat{\varepsilon},\hat{\varepsilon}_\perp)}$ satisfying familiar eigenvalue equations,

$$\left(\hat{a}_{\hat{\varepsilon}}, \hat{a}_{\hat{\varepsilon}_\perp}\right)|\alpha,\beta\rangle_{(\hat{\varepsilon},\hat{\varepsilon}_\perp)} = (\alpha,\beta)\,|\alpha,\beta\rangle_{(\hat{\varepsilon},\hat{\varepsilon}_\perp)}, \tag{14}$$

where $\alpha,\beta$ are classical CAs, we obtain IOP, after inserting eq. (14) into eq. (12), as

$$p_{(\hat{\varepsilon},\hat{\varepsilon}_\perp),n} = \beta^n/\alpha^n. \tag{15}$$

Writing IOP in polar form, $p_{(\hat{\varepsilon},\hat{\varepsilon}_\perp),n} = |p_{(\hat{\varepsilon},\hat{\varepsilon}_\perp),n}| \exp\,(i\Delta_{(\hat{\varepsilon},\hat{\varepsilon}_\perp),n})$ and polar decomposition of classical CAs, $\alpha,\beta$, one may infer that the ratio of real amplitudes in orthogonal basis - modes $(\hat{\boldsymbol{\varepsilon}}, \hat{\boldsymbol{\varepsilon}}_\perp)$ has one non-random value, $|p_{(\hat{\varepsilon},\hat{\varepsilon}_\perp),n}|^{1/n}$, but the phase difference can take any of the 'n' non-random values $\frac{1}{n}(\Delta_{(\hat{\varepsilon},\hat{\varepsilon}_\perp),n} + 2r\pi)$ with r = 0, 1,… (n-1).

Typical examples of HOP are provided by even or odd coherent states [20]. If bimodal optical field is such excited as to possess coherent state in one mode and other mode is in odd or even coherent state, i.e.,if $|\psi\rangle_\pm \equiv |\alpha\rangle_x \frac{1}{\sqrt{2}}[|\beta\rangle_y \pm |-\beta\rangle_y]$, one obtains, $[\hat{a}_y^2 - (\beta^2/\alpha^2)\hat{a}_x^2]|\psi\rangle_\pm = 0$, analogous to eq. (12) but not a relation of the form $[\hat{a}_y - p\hat{a}_x]|\psi\rangle_\pm = 0$. So the optical field, $|\psi\rangle_\pm$ is attributed to second-order HOP in the linear polarization basis $(\hat{\mathbf{e}}_x, \hat{\mathbf{e}}_y)$ with $p_{(\hat{e}_x,\hat{e}_y),2} \equiv \beta^2/\alpha^2$. Instances of still higher HOP states are obtained when the optical fields are in Schrödinger Cat States (21). For, if $|\psi\rangle_\pm \equiv |\alpha\rangle_x \frac{1}{\sqrt{2}}[|\beta\rangle_y \pm |\beta^*\rangle_y]$, where $\beta^n$ is real, one gets $[\hat{a}_y^n - (\beta^n/\alpha^n)\hat{a}_x^n]|\psi\rangle_\pm = 0$ and, therefore, $|\psi\rangle_\pm$ are polarized in the $n^{th}$ order in the basis $(\hat{\mathbf{e}}_x, \hat{\mathbf{e}}_y)$ with IOP, $p_{(\hat{e}_x,\hat{e}_y),n} \equiv \beta^n/\alpha^n$. Furthermore, if both the $\hat{\boldsymbol{\varepsilon}}$ and $\hat{\boldsymbol{\varepsilon}}_\perp$ components are Cat states or Cat-like states, $|\psi\rangle = \left(\sum_{s_1=0}^{n_1-1} c_{s_1}|\alpha e^{i2\pi s_1/n_1}\rangle\,\hat{\boldsymbol{\varepsilon}}\right)\left(\sum_{s_2=0}^{n_2-1} d_{s_2}|\beta e^{i2\pi s_2/n_2}\rangle\,\hat{\boldsymbol{\varepsilon}}_\perp\right)$, it displays $n^{th}$ order polarization, where n being least common multiple of $n_1$ and $n_2$. Realizabilities of HOP may be ascertained by generation of the Cat and Cat-like states [22, 23].

The 'statement of the basis' is more important for HOP than that for usual optical-polarization. For higher ($n^{th}$) order polarized light in one basis may not be polarized in any order in some other basis, or may be polarized in some different order in some different basis. For usual polarized light, the criterion (12) is satisfied with n = 1 for any arbitrarily chosen basis $(\hat{\boldsymbol{\varepsilon}}, \hat{\boldsymbol{\varepsilon}}_\perp)$, although IOP is different for different basis and depends on $\hat{\boldsymbol{\varepsilon}}_0$ (see Eq. 3). As a revelation of the above fact, one may consider that light, being in the pure state $|\psi\rangle = |i\sqrt{3}\,\alpha\rangle_x \frac{1}{\sqrt{2}}[|\alpha\rangle_y + |-\alpha\rangle_y]$, gives $\left(\hat{a}_y^2 + \frac{1}{3}\hat{a}_x^2\right)|\psi\rangle = 0$ showing that light, is polarized in the second order with IOP, p = −1/3 in the basis $(\hat{\mathbf{e}}_x, \hat{\mathbf{e}}_y)$, while in the circular polarization basis $\hat{\mathbf{e}}_\pm = (\hat{\mathbf{e}}_y \pm \hat{\mathbf{e}}_x)/\sqrt{2}$, we can write the same state as, $|\psi\rangle = \frac{1}{2}\left[|\sqrt{2}\,\alpha e^{i\pi/3}\rangle_+|\sqrt{2}\alpha e^{-i\pi/3}\rangle_- + |\sqrt{2}\,\alpha e^{i2\pi/3}\rangle_+|\sqrt{2}\alpha e^{-i2\pi/3}\rangle_-\right]$ giving $(\hat{a}_-^3 - \hat{a}_+^3)|\psi\rangle = 0$ which displays light polarized in $3^{rd}$ order with polarization index 1 in the basis $(\hat{\mathbf{e}}_+, \hat{\mathbf{e}}_-)$.

In conclusion, the concept of Higher-order optical polarization is introduced by generalizing the perfect optical-polarization theory, which is, then, applied to study the polarization properties of light in superposition of coherent states. We hope that HOP finds applications in Quantum Optics.

The authors acknowledge the critical discussions with Professors R. Prakash and N. Chandra.

**References:**

1) M. Born and E. Wolf, *Principles of Optics* (Cambridge University Press, Cambridge, England, 1999)
2) E. Knill, R. Laflamme and G. J. Milburn, "A scheme for Efficient Quantum-computation with Linear Optics," Nature **409,** 46 (2001).
3) P. Kok, W. J. Munro, K. Nemoto, T. C. Ralph, J. P. Dowling and G. J. Milburn, "Linear Optical Quantum-computing with Photonic Qubits," Rev. Mod. Phys **79,** 135 (2007).
4) N. Korolkova, R. Louden, T. C. Ralph and C. Silberhorn, "Polarization Squeezing and Continuous-Variable Polarization Entanglement," Phys. Rev A **65,** 052306 (2002).
5) W. P. Bowen, R. Schnabel, H. A. Bachor and P. K. Lam, "Polarization Squeezing and Continuous-variable Stokes Parameters," Phys. Rev. Lett. **88,** 093601 (2002).
6) H. Prakash and N. Chandra, "Density Operator of Unpolarized Radiation," Phys. Rev. A **4**, 796 (1971).
7) G. S. Agarwal, "On the state of unpolarized radiation," Lett. Nuovo Cim. **1**, 53 (1971).
8) A. Sehat, J. Soderholm, G. Bjork, P. Espinoza, A. B. Klimov and L. L. Sanchez-soto, "Quantum polarization properties of two-mode energy eigen states," Phys. Rev. A **71,**033818 (2005).
9) C. L. Mehta and M. K. Sharma, "Diagonal coherent-state representation for polarized light," Phys. Rev. D **10,** 2396 (1974).
10) H. Prakash and R. S. Singh, "Operator for Optical Polarization," J. Phys. Soc. Japan **69,** 284 (2000).
11) A. Peres, *Quantum Theory: Concept and Methods* (Kluwer Academic Publishers, London, U. K., 2002, pp.09-11).
12) A. Luis, "Quantum Polarization Distribution via Marginals of Quadrature Distributions," Phys. Rev. A **71,** 053801 (2005).
13) V. P. Karassiov, "Polarization Structure of Quadrature Light Fields: a new insight.I. General Outlook," J. Phys. A **26**, 4345 (1993);
14) V. P. Karassiov, "Polarization Squeezing and new States of Light in Qunatum Optics," Phys. Lett. A **190**, 387 (1994).
15) J. Schwinger, "Unitary Operator Bases," Proc. Nat. Acad. Sci. USA **46**, 570 (1960).
16) A. B. Klimov, J. Delgado and L. L. Sanchez-soto "Quantum Phase-Space Description of Light Polarization," Opt. Comm. **258**, 210 (2006).
17) Ravi S. Singh, Sunil P. Singh and Gyaneshwar K. Gupta, "Quasi probability distribution functions of optical polarization," in Photonics 2010, Xth International conference on Fibre Optics and Photonics, S. Khizwania and A. K. Sharma, eds.(Viva Books 2011), pp. 560.
18) L. Mandel and E. Wolf, *Optical Coherence and Quantum Optics* (Cambridge University Press, Cambridge, 1995, pp. 468-471).
19) Using this equation a Polarization Operator $P \equiv \hat{a}_x^{-1}\hat{a}_y$ is obtained [10], where $\hat{a}_x^{-1} \equiv (1+\hat{a}_x^\dagger\hat{a}_x)^{-1}\,\hat{a}_x^\dagger$ is the inverse annihilation operator [C. L. Mehta, A. K. Roy and G. M. Saxena: Phys. Rev. A 46 1565 (1992)] of which action on any perfect optical-polarized state, $|\psi\rangle$ gives a modified eigen value equation furnishing IOP, p as $\hat{P}|\psi\rangle = p\,(1-\hat{V}_x)|\psi\rangle$, $\hat{V}_x$ is the vacuum projection operator for x-linearly polarized virtual photons.
20) V. V. Dodonov, I. A. Malkin and V. I. Manko, "Even and odd coherent states and excitations of a singular oscillator," Physica **72**, 597 (1974).
21) E. Schrödinger, "Die gegenwartige situation in quantenmechanik," Naturwissenschaften **23,** 807 (1935).
22) K. Tara, G. S. Agarwal and S. Chaturvedi, "Production of Schrödinger macroscopic quantum-superposition states in a Kerr medium," Phys. Rev. A **47,** 5024 (1993).
23) M. Dakna, T. Anhut, T. Opatrny, L. Knöll and D.-G. Welsch, "Generating Schrödinger-cat-like states by means of conditional measurements on a beam splitter," Phys Rev A **55**, 3184 (1997).